\documentstyle[12pt]{article}
\def\ra{\rightarrow}

\def\prd#1{{\em Phys. Rev.}~{\bf D#1}\ }

\def\prl#1{{\em Phys. Rev. Lett.}~{\bf #1}\ }
\def\plett#1{{\em Phys. Lett.}~{\bf #1B}\ }
\def\np#1{{\em Nucl. Phys.}~{\bf B#1}\ }
\def\deg{\ifmmode{^{\circ}}\else ${^{\circ}}$\fi}

\def\itm#1{\item[$(#1)$]}
\def\gsim{\,\raisebox{-0.13cm}{$\stackrel{\textstyle>}{\textstyle\sim}$}\,}
\def\lsim{\,\raisebox{-0.13cm}{$\stackrel{\textstyle<}{\textstyle\sim}$}\,}
\def\bi{\begin{itemize}}
\def\ei{\end{itemize}}
\def\ed{\end{document}}
\def\be{\begin{equation}}
\def\ee{\end{equation}}
\def\ba{\begin{array}}
\def\ea{\end{array}}
\def\bea{\begin{eqnarray}}
\def\eea{\end{eqnarray}}
\def\beas{\begin{eqnarray*}}
\def\eeas{\end{eqnarray*}}

\def\req#1{(\ref{eq:#1})}
\def\eq#1{Eq.~(\ref{eq:#1})}
\def\labeq#1{\label{eq:#1}}

\def\tfrac#1#2{{\textstyle\frac{#1}{#2}}}

\def\thalf{\tfrac{1}{2}}
\def\tquarter{\tfrac{1}{4}}
\def\tthird{\tfrac{1}{3}}

\def\ev{\ \mbox{eV}}
\def\gev{\ \mbox{GeV}}
\def\tev{\ \mbox{TeV}}

\def\gsim{\raisebox{-0.5ex}{$\stackrel{>}{\sim}$}}
\def\lsim{\raisebox{-0.5ex}{$\stackrel{<}{\sim}$}}

\def\eb{\end{thebibliography}}

\def\labeq#1{\label{eq:#1}}
\def\req#1{(\ref{eq:#1})}
\def\eq#1{Eq.~(\ref{eq:#1})}
\def\tr{\ifmmode{\mbox{Tr}}\else Tr\fi}
\def\bb{\bibitem}

\def\co#1{${\cal O}(#1)$}

\def\nmt{\nu_{\mu}\mbox{-}\nu_{\tau}}

\def\llb{(\lambda^{\dagger}\lambda)}
\def\gs{g^*}
\def\ms{M_{\rm smaller}}
\def\ml{M_{\rm larger}}
\def\eeo{\epsilon/\epsilon_0}
\def\yl{Y_L}
\def\yb{Y_B}
\def\msl{\left(\frac{\ms}{10^{11}\gev}\right)}
\def\tr{T_{RH}}
\def\trr{(10^{10}\gev/\tr)}

\begin{document}
\begin{titlepage}
\begin{flushright}  {\sl NUB-3205/99-Th}\\
{\sl Sep 1999}\\ hep-ph/9909477
\end{flushright}
\vskip 0.5in

\begin{center}
{\Large\bf Leptogenesis and the Small-Angle MSW Solution}\\ [.5in] {Haim Goldberg}\\ [.1in]
{\it Department of Physics\\ Northeastern University\\ Boston, MA 02115, USA}\\
\end{center}
\vskip 0.4in

\begin{abstract}
The lepton asymmetry created in the out-of-equilibrium decay of a heavy
Majorana neutrino  can generate the cosmological baryon asymmetry $\yb$
when processed through fast anomalous electroweak reactions. In this work I
examine this process under the following assumptions: (1) maximal $\nmt$
mixing (2) hierarchical
mass spectrum $m_3\simeq 5\times 10^{-2}\ev\gg m_2$ (3) small-angle
MSW solution to the solar neutrino deficit. Working in a basis where the
charged lepton and heavy neutrino
mass matrices are diagonal, I find the following bounds on the heavy Majorana
masses $M_i:$
$(a)$ for a symmetric Dirac
neutrino mass matrix (no other constraints),  a $\yb$
compatible with BBN constraints can be obtained for $min(M_2,M_3)>
10^{11}\gev$ $(b)$ if {\em any} of the Dirac
matrix elements vanishes, successful baryogenesis can be effected
for a choice of
$min(M_2,M_3)$ as low as a few $\times 10^{9}\gev.$ The latter is
compatible with reheat requirements for supersymmetric cosmologies with
sub-TeV gravitino masses.
\end{abstract}
\end{titlepage}
\setcounter{page}{2}

\section{Introduction}
The accumulating atmospheric neutrino data from SuperKamiokande \cite{SK}  has
greatly increased the likelihood that neutrinos are massive, and that there is
mixing among the neutrino flavor states. Fits to the zenith angle
distribution are consistent with $(a)$ maximal $\nu_{\mu}$-$\nu_{x}$ mixing and
$(b)$ a
(mass)$^2$ difference between the two mass eigenstates $|\Delta m^2| \simeq
3\times 10^{-3}\ev^2.$ The solar neutrino data, both from SuperK \cite{SKsolar}
and other
experiments \cite{chlorine,sage,gallex}, is as yet less definitive
in constraining the neutrino masses and
mixing: there exist the small-angle  and large-angle  solutions \cite{mswsol}
of the MSW effect\cite{msw}, as
well as the vacuum oscillation solution. Omitting sterile neutrinos from
consideration (perhaps with some unwarranted prejudice), one finds that each of
these is not yet a clear favorite: the day-night asymmetry of solar neutrinos,
if persistent
at  higher statistical significance, would disfavor the small-angle MSW and the VO
solutions, while the reported recoil electron energy spectrum at SuperK requires
significant experimental or theoretical modification at larger recoil energies
in order to be compatible with the large-angle solution. It is safe to say that as yet
none of the three is ruled out.

These recent advances have renewed interest in leptogenesis as the precursor to
the establishment of the cosmological baryon asymmetry $\yb\equiv
n_B/g^*n_{\gamma}\simeq 0.6 - 1.0\times 10^{-10}$ required for a successful
description of nucleosynthesis \cite{BBN}. (Here $g^*$ is the effective number
of spin degrees of freedom.) In the simplest leptogenesis scenario \cite{fky},
which forms the basis for the discussion in this paper, a $B-L$ asymmetry is
established through the $CP$- and $L$-violating out-of-equilibrium decay of the
neutral heavy Majorana lepton which partakes in the see-saw mechanism
\cite{seesaw} for the light neutrino masses. (The atmospheric oscillation data
indicating very small mass differences can be taken as supportive of the
see-saw mechanism.) In the next stage of this scenario, the $B-L$ asymmetry
is reprocessed through the fast
$(B+L)$-violating anomalous processes \cite{sph,spha} preceding the electroweak
phase transition into the required $\yb$ \cite{harvturn}.
Because the successful completion of
this process places non-trivial constraints on the both the Dirac and Majorana
sectors of the neutrino mass matrix, it has been extensively discussed in this context
in the
literature \cite{everyone}. The approach commonly taken is to explore the
implications for leptogenesis of various models or {\em ansatze} for the three
relevant mass matrices: $m_D\equiv m_{\nu N}^{\rm Dirac}, m_{\ell}\equiv
m_{\ell\ell^c}^{\rm Dirac},
\ \mbox{and}\ M\equiv M_{NN}^{\rm Majorana}.$ ($\ell$ is the charged lepton.)
These constitute important exercises in relating the neutrino data and the
baryon asymmetry to the underlying Yukawa structures. However, it turns out
that in certain experimentally allowed regions of the light neutrino
mass matrix the analysis becomes
much less dependent on the details of the Yukawa matrices, and compliance with
the BBN-consistent
$\yb$ can translate more directly to a  constraint on the heavy Majorana
masses. This is the work of this paper: I will find that
for a hierarchical light
neutrino spectrum, maximal
$\nmt$ mixing, and the small-angle MSW solution to the solar neutrino deficit,
agreement with the cosmological $\yb$ can generally obtained for a (lightest) Majorana
mass as small as
$10^{11}$ GeV. Moreover, when any of the matrix elements of $m_D$ are zero
(in a basis where $m_{\ell}
\ \mbox{and}\ M$ are diagonal), this bound can be lowered to $\sim 10^9$\gev.
In the concluding section I discuss the implication of these numbers for
the gravitino problem \cite{reheat} of supersymmetric cosmology.

\section{Assumptions}
In what follows, I will work under the following assumptions consistent with
present  neutrino data:
\bi\itm{a} Light neutrino mass hierarchy $|m_3|\simeq 5\times 10^{-2}\ \ev\gg |m_2|\gg m_1$
\itm{b} Maximal $\nmt$ mixing, consistent with SuperK
atmospheric data \cite{weiler,goldhaber}
\itm{c} Small-angle MSW  solution for the solar neutrino deficit
\itm{d} (Theory) The seesaw mechanism \cite{seesaw} is operative, with a
hierarchical structure in the three heavy Majorana masses.
\ei
The hierarchical assumption $(a)$ rules out  consideration of a
nearly-degenerate scenario for neutrino masses in the 2-3 sector.
Assumption $(c)$ (the adoption of the small
angle MSW solution) is not dictated by observation. Compared to the large-angle
MSW, it provides a marginally better (but not good) fit to the recoil electron
energy spectrum at SuperK, and a less good fit to the day-night variations
\cite{mswsol}. More data, and perhaps a better understanding of the {\em hep}
neutrino spectrum, will decide the issue. However, it is assumption $(c)$ which
allows the study of the constrained system, and that is the reason for its
adoption \cite{laz}. These assumptions also allow one to ignore renormalization
group effects in running from the Majorana to low energy scale \cite{rg}.

\section{Seesaw Relation for Small Angle MSW}
Under Assumption $(b),$ the electron neutrino plays no role in the atmospheric
neutrino anomaly, and the light neutrino mixing matrix is given by
\cite{weiler,goldhaber}

\be U=\left(\ba{ccc}
\cos\theta & \sin\theta & 0\\
-\sin\theta/\sqrt{2}& \cos\theta/\sqrt{2}&1/\sqrt{2}\\
\sin\theta/\sqrt{2}& -\cos\theta/\sqrt{2}& 1/\sqrt{2}
\ea\right)\ \ ,
\labeq{u}
\ee
where
\be U^T\ m_{\nu}\ U = diag\ (e^{i\phi_1}\ m_1,\; e^{i\phi_2}\; m_2,\ m_3)\ \ .
\labeq{umu}
\ee
The phases ($\phi_1,\phi_2)$ are Majorana $CP$-violating phases, and I have
omitted the $CP$-violating CKM-type phase in $U.$ The solar mixing angle is
given by $\sin^2 2\theta,$ and for the small-angle MSW solution $\sin^2
2\theta\simeq 5\times 10^{-3}$ \cite{mswsol}, so that $\theta\simeq 0.035.$

The seesaw mechanism is expressed by
\bea
m_{\nu}&=&m_D\ M^{-1}m_D^{\;T}\\
&=& v^2 \lambda\ M^{-1}\lambda^T
\labeq{seesaw}
\eea
where the matrices $\lambda$ and $M$ are defined through the Lagrangian
\be
{\cal L}= - L\lambda NH+ \thalf NMN\ \ ,
\labeq{lagr}
\ee
so that $m_D=\lambda v,$ $v=174\gev.$ $\lambda$ is a $3\times 3$ complex
Yukawa matrix. I will work in a basis where the charged lepton masses and $M$
are diagonal, so that $M^{-1}=diag\ (1/M_1,\ 1/M_2,\ 1/M_3).$ Even if we were to
know all six independent elements of the symmetric matrix $m_{\nu}$ as well as
the values of $M_1,\ M_2,\ M_3,$ the seesaw condition \req{seesaw} provides six
equations for the nine complex unknown matrix elements of $\lambda.$ In general,
the leptogenesis scenario requires knowledge of the entire matrix $\lambda,$ so
that considerable input besides \req{seesaw} is needed
in order to determine
$\lambda.$

The situation improves a great deal for the small-angle MSW solution. I will
simplify matters by taking
$\theta=0$ in \eq{u}, and $m_1=0, M_1\gg M_{2,3}$. These are sufficient to decouple
$\nu_e$ from the $\nmt$ seesaw and from the leptogenesis
scenario. On input of Eqs.\req{u} and \req{umu}, the seesaw
equation \req{seesaw} (now $2\times 2)$ reduces to a set of three equations for the four
(complex) unknowns $a, b, c, d,$ if we regard
$m_2, m_3, M_2, M_3$ as `known':
\bea
(a^2-c^2)/M_2 + (d^2-b^2)/M_3&=&0\nonumber\\
a(a-c)/M_2 + d(d-b)/M_3&=&m_2/v^2\nonumber\\
a(a+c)/M_2+d(d+b)/M_3&=&m_3/v^2\ \ .
\labeq{ssawthree}
\eea
The Dirac Yukawa matrix $\lambda$ has been parameterized as
\be\lambda=\left(\ba{cc}a&d\\c&b\ea\right)\ \
\labeq{ldef}
\ee
and the Majorana phase $\phi_2\equiv \phi$ is incorporated into $m_2:\
m_2=|m_2|e^{i\phi}.$
\section{The Leptogenesis Scenario}
The leptogenesis scenario has been carefully discussed by many authors \cite{lepref}.
Briefly, the present baryon asymmetry of the universe is calculated in the
following manner:
\bi \itm{a} First, the lepton asymmetry $\yl$ is given in terms of the decay
asymmetry $\epsilon$ of $N_i,$ the lightest of the $N$'s, parameterized as
follows:
\be \yl=\frac{n_L-n_{\bar L}}{\gs n_{\gamma}}=\kappa \ B \ \epsilon\ \ ,
\labeq{yl}
\ee
where
\be
\epsilon \simeq \frac{3}{16\pi}\frac{1}{\llb_{ii}}\ \sum_{j\ne i}\ Im
\left[\llb_{ij}^{\;2}\right]\ \frac{M_i}{M_j}\\
\labeq{epsi}
\ee
on the assumption of a mass hierarchy $M_i\ll M_{j\ne i}$ \cite{fky,epsform}.
The meanings of $\kappa$ and $B$ are as follows:
\bi
\itm{i}{\em Thermal production:} If $M_{\rm smaller}\equiv min(M_2,M_3)< \tr$
($\tr$= post-inflation reheat temperature)
and the inverse decay rate is sufficient to establish
equilibrium, then
\bea
\kappa&=&\mbox{suppression factor due to washout by
inverse decay}\nonumber\\[-1mm]
& & \mbox{and}\ 2\ra 2\ \mbox{lepton-violating scattering
processes}\nonumber\\
B&=&1/\gs
\labeq{inequ}
\eea
where $\gs$ is the effective number of massless spin degrees of freedom at the
time of $N$ decay $(\gs=106.75$ in the Standard Model). The factor $\kappa$ is
determined by numerical integration of the Boltzmann equations
\cite{BBN,boltzmann} and depends most sensitively on the ratio
\bea
K&\equiv&\Gamma_i/H(T=M_i)\nonumber\\
&=&\left(\frac{\llb_{ii}M_i}{8\pi}\right)\ \left(\frac{1.7\sqrt{\gs} M_i^2}{M_{{\rm
P}\ell}}\right)^{-1}
\labeq{k}
\eea
The suppression factor $\kappa$ reaches its
limiting value of 1.0 for $K\ll 1,$
and drops to $\simeq 0.01$
for $K=20.$
\itm{ii}{\em Non-thermal production via inflaton decay:}
If $\ms/\tr\gsim 10$ and $1\lsim K\lsim 100$, then integration of the Boltzmann equations
(starting at $M/T\ge 10)$ reveals negligible suppression due to inverse decays, and
\cite{outofeq}
\bea
 \kappa &\simeq& \tr/M_{inflaton}\simeq 10^{-3}(\tr/10^{10}\gev)\nonumber\\
 B&=& \mbox{average number of $N$'s produced in decay of an inflaton} .
 \labeq{outofeq}
\eea
\ei
\itm{b}Finally, the baryon asymmetry is established when the $B-L$ asymmetry
is processed through the fast $(B+L)$-conserving sphaleron processes \cite{sph,spha}
above the
electroweak transition temperature, and is given by \cite{ltob}
\be
\yb\simeq -\tthird\  \yl\ \ .
\labeq{yb}
\ee\ei

\section{Results}
It is clear from \req{epsi} that a calculation of $\yb$ will involve all the
matrix elements of $\lambda,$ so that even the truncated $2\times 2$ seesaw
equations are not quite sufficient to enable a casting of $\yb$ in terms of the
masses alone. I will give results for the following illustrative constraints on
$\lambda$:
\bi\itm{1}$a,\ b,\ c,\ d,$ respectively, are set $=0.$
\itm{2}$c=d$ (symmetric case)
\itm{3}$a=b$\ei
Each of these will be worked out for both cases $(i)\ M_2\gg M_3$ $(ii)\ M_3\gg
M_2.$ In all cases, the hierarchy $m_3\gg |m_2|$ will be respected.

The results listed in Table I are obtained by inserting these constraints
into \eq{seesaw}, solving
for the matrix elements, and then utilizing Eqs.\ \req{epsi} and \req{k} to
calculate $\epsilon$ and the out-of-equilibrium parameter $K.$ The quantities
$x,\ y,\ K_0,\ \mbox{and}\ \epsilon_0$ in the Table are defined as follows:
\bea
x&\equiv& |m_2|/m_3\nonumber\\[2mm]
y&\equiv& \ms/\ml\nonumber\\[2mm]
K_0&\equiv&m_3/m_0\ \ ,\quad m_0=(1.7)8\pi\sqrt{\gs}v^2/M_{{\rm P}\ell}\simeq
1.1\times 10^{-3}\ \ev\nonumber\\[2mm]
&\simeq& 45\nonumber\\[2mm]
\epsilon_0&\equiv&\frac{3}{16\pi}\frac{m_3\ms}{v^2}
\simeq 10^{-5}\left(\frac{m_3}{0.05\ \ev}\right)\msl .
\labeq{defs}
\eea\medskip

\[\begin{array}{|c|c|c|c|c|}\hline
\ms=M_3 & \ms=M_2 & K/K_0 & \eeo & \kappa\ (\ms<\tr) \\ \hline\hline
a=0\ \mbox{or}\ c=0 & b=0\ \mbox{or}\ d=0 & 1 & x \sin\phi & 4.0\times 10^{-3}\\
\hline
b=0\ \mbox{or}\ d=0 & a=0\ \mbox{or}\ c=0 & 2x &  \thalf \sin\phi &
1.0\times 10^{-2}\ (x=0.25)\\ \hline
c=d\ \mbox{or}\ a=b & c=d\ \mbox{or}\ a=b & 1 &
\left(4xy\right)^{1/2}\sin\thalf\phi & 4.0\times 10^{-3}\\ \hline
\end{array}
\]
\noindent\hspace*{0.6in}Table I:~ \parbox[t]{4in}{Results for
illustrative constraints. For clarity of presentation, various overall $\pm$
signs have been omitted.} \vskip .5in

\noindent Two comments with respect to the results in Table I are in order:
\bi \item
In obtaining the result for $\eeo$ in Line 3, one finds that the algebra simplifies greatly
if  $x\ll \tquarter y.$ The result given reflects this choice.
\item As noted above, in the case of thermal production,
the suppression factor $\kappa$ depends on
$K,$ and was obtained by integration of the rate equations
\cite{BBN,boltzmann}, subject to the initial conditions $Y_N(M/T\simeq 0)=1/\gs,$
$Y_L(M/T\simeq 0),$ where $Y_N\equiv n_N/\gs n_{\gamma}.$ For non-thermal
production, $\kappa$ is given in \eq{outofeq} above. \ei
I now proceed to calculate $\yb$ and require
\be
\yb=\yb^{\rm BBN}\ge 0.6\times 10^{-10}\ \ .
\labeq{ybbn}
\ee
{}From Eqs. \req{yl},\req{yb}, and
\req{defs}, one obtains (ignoring signs)
\be
\yb=3\times 10^{-8}\ \kappa\ (B\gs) (\eeo)\ \msl\
\ ,
\labeq{yba}
\ee
where I have taken $m_3=5.0\times 10^{-2}\ \ev.$ From \req{yba} and \req{ybbn},
there results a lower bound on the lighter of $(M_2, M_3)$:
\be
\kappa\ (B\gs)\ (\eeo)\ M_{\rm smaller}\ge 2\times 10^8\ \gev\ \ .
\labeq{lower}
\ee

Incorporating the requirements $x\le \tquarter$ (mass hierarchy), $|\sin\phi|\le 1,$
the three lines in Table I can be addressed in turn for each of the scenarios,
and a bound obtained from \eq{lower}:
\medskip

\noindent{\bf Thermal Production:}\medskip

\noindent{\em Line 1:}
\be
\ms\ge 2\times 10^{11}\ \gev\ \ .
\labeq{mseqa}\ee

\noindent{\em Line 2:} In this case $\kappa$
depends strongly on $x$ through the dependence of $K=2xK_0,$ and this is
reflected in the bound for $\ms.$
\bea
x=0.25:\quad \ms&\ge& 4\times 10^{10} \gev\nonumber\\
x=0.04:\quad \ms&\ge& 4\times 10^9\ \gev
\labeq{mseqb}
\eea
\noindent{\em Line 3:} As noted above, the expression for $\eeo$ given in the
Table reflects a simplifying constraint $x\ll \tquarter y.$ The resulting bound is
\be
\ms \ge 5 y^{-1}\times 10^{10}\gev\ge 2\times 10^{11}\gev
\labeq{mseqc}
\ee
if we take $y < \tquarter$ in order to maintain the hierarchy in the heavy
masses.
\vskip 0.3in

\noindent{\bf Non-Thermal Production:} Here $\kappa$ is given
by \eq{outofeq}. In terms of a scaling factor
\[
\zeta \equiv \trr \ (100B)^{-1}\,
\]
one finds

\noindent{\em Line 1:} For $x\le\tquarter$
\be
\ms\ge 8\times 10^{11}\ \zeta\ \gev
\labeq{msnequa}
\ee

\noindent{\em Line 2:}

\be
\ms\ge 4\times 10^{11} \ \zeta\ \gev
\labeq{msnequb}
\ee

\noindent{\em Line 3:} With the same restrictions as in the previous section,
\be
\ms\ge 1\times 10^{12}\ \zeta\ \gev\ \ .
\labeq{msnequc}
\ee

\section{Discussion of Results and Conclusions}
\noindent(1)~This work has focused on a particular sector of the neutrino
mass spectrum (hierarchical) and mixing matrix (maximal $\nmt$ mixing, small-angle
MSW). With the seesaw mechanism, this greatly constrains the Dirac
Yukawa matrix $\lambda$, effectively decoupling the electron neutrino and one of the heavy
Majoranas.
As a consequence, a single additional condition on the four complex matrix elements
of the effective $\lambda$ allows its determination in terms of light and heavy masses,
and a single CP-violating Majorana phase. The lepton asymmetry
resulting from out-of-equilibrium decays of the heavy Majorana neutrinos may
then be computed in terms of these parameters.
Comparing the  resulting cosmological baryon asymmetry with the value required
{}from BBN then turns out to place a lower bound on the lighter Majorana
mass.\medskip

\noindent(2)~In the case of {\em thermal production},
for the range of scenarios studied (including several not reported in this
paper, such as $b=c$), the lower bounds found for the mass of the lightest
heavy Majorana are typically of \co{10^{11}}\gev, well below the inflaton mass
of $\sim 10^{13}\gev.$ Thus, the heavy Majorana may be produced during
reheating via inflaton decay, without recourse to parametric resonance
production \cite{param}. A reheat temperature of \co{10^{11}}\gev\ requires a
large gravitino mass $\gsim 2.5$\tev\ \cite{grav} in order that decays of
produced gravitinos not destroy the products of nucleosynthesis. If one of the
entries in the Dirac Yukawa is zero, there are scenarios (Line 2 of Table I),
in which the smaller Majorana mass may fall below $10^{10}\gev,$ which is a
safe reheat temperature for low gravitino masses. In the case of {\em
non-thermal production}, the lower bounds are a bit higher, and can exceed the
inflaton mass $\sim 10^{13}\gev$ if the reheat temperature is less than
$10^9\gev$ (see Eqs.
\req{msnequa}-\req{msnequc}). In that case, production via parametric resonance
would be necessary.\medskip

\noindent (3)~Various small parameters, such as the solar mixing angle $\theta$
or the $U_{e3}$ element of the mixing matrix, have been set to zero. In
principle, small entries for these can compete with the mass hierarchy
parameters $x$ and $y,$ and cloud the results of this work \cite{altarelli}. As
a crude measure, one can limit the present discussion to values of
$x,y\ \gsim\ \theta\simeq 0.03.$ However, for the present CHOOZ bound
$|U_{e3}|^2\le 5\times 10^{-2},$ a similar criterion
$x,y\ \gsim \ |U_{e3}|_{max}\simeq 0.2$ may be too restrictive. As the data improves,
the effects of any small non-zero entries can be assessed.\medskip

\noindent(4)~The discussion presented here
is considerably more constrained than previous studies \cite{everyone} which
assume entire textures
for both the Dirac and Majorana matrices, and often
leave undetermined a good number of parameters.
As stated in the introduction, such studies are valuable as links to larger
theories of flavor symmetries, and are more flexible in accommodating a changing
scenario for the neutrino parameters. The aim here is much more
phenomenological,
incorporating {\em ab initio}
certain constraints on the light neutrino mass matrix, and leaving to vary only
one complex parameter. Of course, increased statistics
on the day-night effect could
begin to seriously disfavor the small-angle MSW solution, thus removing the basis
for the simplification in this work.

\section{Acknowledgments}

I would like to thank the organizers of the Low-Energy Neutrino Workshop,
Institute for Nuclear Theory, University of Washington, Summer 1999, where this
work was begun. Thanks also to Hitoshi Murayama and Jose Valle for enlightening
conversations. This research was supported in part by the National Science
Foundation through Grant No. PHY-9722044.

\begin{thebibliography}{99}
\bb{SK}SuperKamiokande Collaboration, Y. Fukuda {\em et al}, \prl{81}(1998)
1562; \prl{82}(1998) 2611.
\bb{SKsolar}SuperKamiokande Collaboration, Y. Fukuda {\em et al}, \prl{81}(1998)
1158; (Erratum-{\rm ibid}. (1998) 4279; \prl{82}(1999) 2430.
\bb{chlorine}B.T. Cleveland {\em et al.,} Astrophys. J.{\bf
496} (1998) 505; B.T. Cleveland {\em et al.,}\np{38}(Proc. Suppl) (1995) 47.
\bb{sage}SAGE Collaboration (J.N. Abdurashitov {\em et al.,}
astro-ph/9907031 and references therein.
\bb{gallex}GALLEX Collaboration (W. Hampel {\em et al.,} \plett{388}(1996) 364;
\plett{447}(1999)127.
\bb{mswsol}For some recent analyses, see
J.N. Bahcall , P.I. Krastev, and A.Yu. Smirnov, hep-ph/9905220;
M.C. Gonzalez-Garcia, P.C. de Holanda, C. Pena-Garay, and J.W.F. Valle,
hep-ph/9906469; J.N. Bahcall, P.I. Krastev, and A.Yu. Smirnov
\prd{58}(1998) 096016; M. Smy, SuperKamiokande Collaboration, hep-ex/9903034.
See also N. Hata and P. Langacker, \prd{56}(1997) 6107; G.L. Fogli, E. Lisi,
and D. Montanino, {\em Astropart. Phys.} {\bf 9} (1998) 119.
\bb{msw}L. Wolfenstein, \prd{17}(1978) 2369; S.P. Micheyev and A.Yu. Smirnov,
{\em Yad. Fiz.} {\bf 42}(1985) 1441 [{\em Sov.J.Nucl.Phys.} {\bf 42}(1985) 913].
\bb{BBN}E.W. Kolb and M.S. Turner, {\em The Early Universe},
(Addison-Wesley, Reading, MA, 1989).
\bb{fky}M. Fukujita and T. Yanagida, \plett{174}(1986) 45.
\bb{seesaw}M. Gell-Mann, P. Ramond, and R. Slansky, in
{\em Supergravity},  eds. P. Van Nieuwenhuizen and D. Freedman (North-Holland,
Amsterdam, 1979); T. Yanagida, in {\em Proceedings of the Workshop on Unified
Theories and Baryon Number in the Universe},  eds. A. Sawada and A. Sugamoto,
(KEK Report No. 79-18, 1979).
\bb{sph}V.A. Kuzmin, V.A. Rubakov, and M.E. Shaposhnikov, \plett{155}(1985) 36.
\bb{spha}P. Arnold and L. McLerran, \prd{36}(1987) 581.
\bb{harvturn}S.Y. Khlebnikov and M.E. Shaposhnikov, \np{308}(1988) 885;
J.A. Harvey and M.S. Turner, \prd{42}(1990) 3344.
\bb{everyone}M.A. Luty, \prd{45}(1992) 455;
K. Enqvist and I. Vilja, \plett{299}(1993) 281;
T. Gherghetta and G. Jungman, \prd{48}(1993) 1546;
M.P. Worah, \prd{53}(1996) 3902;
W. Buchm\"uller and M. Plu\"macher, \plett{389}(1996) 73;
M. Pl\"umacher, {\em Z. Phys.}~{\bf C74} (1997) 549; \np{530}(1998) 207;
G. Lazarides, Q. Shafi, and N.D. Vlachos, \plett{427}(1998) 53;
W. Buchm\"uller and T. Yanagida, hep-ph/9810308;
J. Ellis, S. Lola, and D.V. Nanopoulos, hep-ph/9902364;
A.S. Joshipura and E.A. Paschos, hep-ph/9906498;
K. Kang, S.K. Kang, and U. Sarkar, hep-ph/9906486;
M.S. Berger, hep-ph/990649;
M.S. Berger and B. Brahmachari, hep-ph/9903406.
\bb{reheat}J. Ellis, J.E. Kim, and D.V. Nanopopulos, \plett{145}(1984) 181;
M. Y. Khlopov and A. D. Linde, \plett{138}(1984) 265. For a recent analysis,
see E. Holtmann, M. Kawasaki, K. Kohri, and T. Moroi, \prd{60}(1999) 023506.
\bb{weiler}V. Barger, S. Pakvasa, T.J. Weiler, and K. Whisnant,
\plett{437}(1998) 107.
\bb{goldhaber}A.J. Baltz, A.S. Goldhaber and M. Goldhaber, \prl{81}(1998)
5730.
\bb{laz}For another analysis based on the small-angle solution, see G.
Lazarides {\em et al.,} ref. \cite{everyone};
G. Lazarides and N. D. Vlachos, \plett{459}(1999) 482; G. Lazarides, hep-ph/9905450.
\bb{rg}K.S. Babu, C.N. Leung and J. Pantaleone, \plett{319}(1993) 191.
\bb{lepref}References in addition to those in \cite{everyone} and
\cite{epsform} may be be found in A. Pilaftsis, {\em Int. J. Mod. Phys.}
{\bf A14} (1999) 1811. Leptogenesis through an Affleck-Dine-type mechanism has
been studied by H.~Murayama and T.~Yanagida, \plett{322}(1994) 349;
T. Moroi and H. Murayama, hep-ph/9908223.
\bb{epsform}M. Pl\"umacher, ref. \cite{everyone};
L. Covi, E. Roulet, and F. Vissani, \plett{334}(1996) 169;
J. Liu and G. Segr\'e,\prd{48}(1993) 4609;
\bb{boltzmann}M. A. Luty, ref. \cite{everyone};
M. Flanz, E.A. Paschos and U. Sarkar, \plett{345}(1995) 248;
M. Flanz, E.A. Paschos, U. Sarkar and J. Weiss, \plett{389}(1996) 693;
M. Pl\"umacher, ref. \cite{everyone}; J. Faridani, S. Lola, P.J. O'Donnell and
U. Sarkar, hep-ph/9804261;
M. Flanz and E.A. Paschos, hep-ph/9805427; J. Ellis {\em et al,} ref. \cite{everyone}.
\bb{outofeq}B.A. Campbell, S. Davidson and K.A. Olive, \np{399}(1993) 111; G. Lazarides
{\em et al.,} ref. \cite{everyone}; G. Lazarides, hep-ph/9904428;
G.F. Giudice, M. Peloso, A. Riotto and I. Tkachev, hep-ph/9905242;
T. Asaka, K. Hamaguchi, M. Kawasaki, and T. Yanagida, hep-ph/9906366.
\bb{ltob}S.Y. Khlebnikov and M.E. Shaposhnikov, \np{308}(1988) 885;
J.A. Harvey and M.S. Turner, \prd{42}(1990) 3344.
\bb{param}G.F. Giudice {\em et al.,} ref. \cite{outofeq}.
\bb{grav}E. Holtmann {\em et al.,} ref. \cite{reheat}.
\bb{altarelli}G. Altarelli and F. Feruglio, hep-ph/9905536.
\eb
\ed